\documentclass[
reprint,
superscriptaddress,
amsmath,amssymb,
aps,
prl,
]{revtex4-1}

\usepackage{graphicx,epsf}
\usepackage{float}
\usepackage{xcolor}
\usepackage{bm}

\begin{document}

\title{Conductance oscillations and speed of chiral Majorana mode
in a quantum-anomalous-Hall 2d strip}

%
\author{Javier Osca}
\affiliation{Institute of Interdisciplinary Physics and Complex Systems IFISC 
(CSIC-UIB), Palma de Mallorca, E-07122, Spain} 
\author{Lloren\c{c} Serra}
\affiliation{Institute of Interdisciplinary Physics and Complex Systems IFISC 
(CSIC-UIB), Palma de Mallorca, E-07122, Spain} 
\affiliation{Department of Physics, University of the Balearic Islands, 
Palma de Mallorca, E-07122, Spain}

\begin{abstract}
We predict conductance oscillations  in a quantum-anomalous Hall 2d strip
having a superconducting region of length $L_x$ with a chiral Majorana mode. These oscillations require a finite transverse
extension of the strip $L_y$ of a few microns or less. Measuring the conductance periodicity with $L_x$ 
and a fixed bias,
or with bias and a fixed $L_x$, yields the speed of the chiral Majorana mode. The physical mechanism 
behind the oscillations is the interference between  backscattered chiral modes from second to first interface of the NSN double junction. The interferometer effect is enhanced by the presence of side barriers. 
\end{abstract}

\maketitle

Majorana modes in Condensed Matter systems are object of an intense research due to their 
peculiar exchange statistics that could allow implementing a robust quantum computer \cite{Nayak}.
A breakthrough in the field was the observation of zero-bias anomalies in semiconductor nanowires 
with proximity-induced superconductivity \cite{Mourik}. Although an intense theoretical debate has followed the experiment, 
evidence is now accumulating \cite{HaoZ,HaoZ2} that these quasi-1d systems indeed host localized Majorana
states on its two ends, for a proper choice of all the system parameters 
(see Refs.\ \cite{Aguado,Lutrev} for recent reviews).

Another breakthrough is the recent observation of a peculiar conductance quantization, $0.5e^2/h$, in a quantum-anomalous-Hall 2d (thin) strip, of $2\times1$ mm \cite{He294}. A region of $L_x\approx 0.8\, {\rm mm}$ in the central part of the strip is put in proximity of a superconductor bar, whose influence 
makes the central piece of the strip become a topological system able to host a single chiral Majorana mode.
The device can then be seen as a generic NSN double junction with tunable topological properties. 
The hallmark of transport by a single chiral Majorana mode in the central part is the observed halved quantized conductance,
since a Majorana Fermion is half an electron and half a hole \cite{Qi10,Chung,Wang13,Wang14,Wang15,Lian16}.

The observed signal of a chiral Majorana mode in the macroscopic device of Ref.\ \cite{He294} naturally leads 
to the question of how is this result affected when the system dimensions are reduced and quantum properties
are enhanced. Can the presence of a Majorana mode still be clearly identified in a smaller strip? 
Are there additional {\em smoking-gun} signals?  In this Rapid Communication we provide theoretical evidence
predicting a positive answer to these questions.  In a smaller strip, with lateral extension $L_y$
of a few microns or less, we predict the existence of conductance oscillations as a function of
$L_x$ (the longitudinal extension of the superconducting piece) and a fixed  
longitudinal bias. Alternatively,  oscillations are also present as a function of bias and a fixed value of
$L_x$. 

We find that the period of the conductance oscillations is related to the speed $c$ of the chiral Majorana mode, as defined
from the linear dispersion relation $E=c \hbar k$  with $k$ the mode wavenumber. More precisely, the oscillating part of the conductance is $\approx\delta{G}\, \cos{(2L_x E/\hbar c)}$.
Therefore, measuring the distance $\Delta L_x$ between two successive conductance maxima in a fixed bias $V$ (equivalent to a fixed energy $E$ since $E=eV/2$), it is $c=\Delta L_x E/\pi\hbar$. Analogously, in a device with a fixed $L_x$ the distance $\Delta E$ between two successive bias maxima yields $c= L_x\Delta E/\pi\hbar$.
As anticipated, this result implies the possibility of a purely electrical measurement of the speed of a 
chiral Majorana mode and it thus provides an additional hallmark of the presence of such peculiar modes.

A narrow strip  (typically $L_y\lesssim 1\,\mu{\rm m}$) is required for a sizable oscillation amplitude  $\delta G$ for, otherwise, it becomes negligible when $L_y$ increases. The amplitude is also weakly dependent on the energy, yielding a slightly damped oscillation with bias. The physical mechanism 
behind the predicted conductance oscillations is the interference of the two backscattered chiral modes
of the NSN double junction. An initially reflected mode  from the  first
interface (NS),
assuming left to right incidence,
is superposed by the reflection from the second interface (SN) that has travelled
backwards a distance $L_x$. A constructive interference yields an enhanced Andreev reflection 
that, in turn, yields an enhanced device conductance. Since Majorana-mode backscattering 
between the two interfaces can be seen as
a manifestation of the hybridization of the two edge modes of the strip, a finite (small) $L_y$
is required for its manifestation as a sizeable effect. 

Chiral mode backscattering in presence of Majorana modes
was already considered in Ref.\ \cite{Chung}, but only between interfaces and leads, not in between interfaces, which only occurs in the quantum limit for smaller $L_y$. We stress that, as discussed in 
Ref.\ \cite{Chung}, we assume a grounded superconductor configuration and that conductance is measured with an applied  symmetrical bias on both sides.
Very recent works \cite{Xie,Zhou18} have considered biased superconductor configurations, focussing
on the dependence of the quasiparticle-reversal transmissions with the chemical potential.
Quasiparticle reversal was also discussed with quantum spin Hall insulators
in Ref. \cite{Rei13}.
Interferometry with Majorana beam splitters was considered in \cite{Akh09,Fu09,Roi18},
although a conceptual difference with our work is that beam splitters constrain quasiparticles to follow different trajectories while we consider a single material strip.
It is also worth stressing that
the suggested interference of this work requires travelling modes 
with a well defined $k$, such as
chiral Majorana modes, since the oscillations persist for arbitrary large values of $L_x$.
This is an essential difference with hybridization of Majorana  (localized) end states in nanowires,
lacking a well defined $k$,
that has been shown to  rapidly vanish as $L_x$ increases \cite{DasS12}.  

{\em Model and parameter values.-}
We consider a model  of a double-layer QAH strip with induced superconductivity 
in the central region as in Ref. \cite{He294}. 
Using vectors of Pauli matrices for the two-valued variables representing usual spin 
$\bm{\sigma}$, electron-hole isospin $\bm{\tau}$ and layer-index pseudospin $\bm{\lambda}$, 
in a Nambu spinorial representation that groups 
the field operators in the top $(t)$ and bottom $(b)$ layers,
$\left[
(
\Psi^t_{k\uparrow},
\Psi^t_{k\downarrow},
\Psi^{t\dagger}_{-k\downarrow},
-\Psi^{t\dagger}_{-k\uparrow}
),
(
\Psi^b_{k\uparrow},
\Psi^b_{k\downarrow},
\Psi^{b\dagger}_{-k\downarrow},
-\Psi^{b\dagger}_{-k\uparrow}
)\right]^T
$,
the Hamiltonian reads
\begin{eqnarray}
{\cal H} &=& 
\left[\, m_0 + m_1 \left(p_x^2 +p_y^2\right)\, \right] \tau_z\, \lambda_x 
+ \Delta_Z\, \sigma_z \nonumber\\
&-& \frac{\alpha}{\hbar}\, \left(\,p_x\sigma_y-p_y\sigma_x\right)\, \tau_z\,\lambda_z \nonumber\\
&+& \Delta_p\, \tau_x + 
\Delta_m\, \tau_x\,\lambda_z\, .
\end{eqnarray}

The physical origin of the different parameters has been discussed in Refs.\ \cite{He294,Qi10,Chung,Wang13,Wang14,Wang15,Lian16}.
Let us only emphasize here that the superconductivity parameters $\Delta_{p,m}$ 
are just the half-sum or half-difference of the corresponding  parameter in each layer, 
$\Delta_{p,m}\equiv (\Delta_1\pm\Delta_2)/2$, with $\Delta_{1,2}$ vanishing in the left and right 
normal regions and taking constant values for $x\in[-L_x/2,L_x/2]$ (see sketch in Fig.1). The strip confinement along the lateral coordinate ($y$)
is obtained by assuming that $m_0$ takes a large value for $y \not\in [-L_y/2,L_y/2]$, effectively 
forcing the wave functions to vanish at the lateral edges.

It is important to consider realistic values for the Hamiltonian parameters. In our calculations we assume 
a unit system set by 1 meV as energy unit, 1 $\mu$m as length unit, and a mass unit $m_U$ from the 
condition 
$\hbar\equiv  m_U^{1/2} {\rm meV}^{1/2} {\mu}{\rm m}$, yielding $m_U=7.6\times 10^{-5} m_e$, 
where $m_e$ is the bare electron mass.
We have assumed $\alpha = 0.26\, {\rm meV}\mu{\rm m}$, $m_0=1\, {\rm meV}$,
$m_1=10^{-3}\,m_U^{-1}$,  $\Delta_{1}= 1\, {\rm meV}$ and $\Delta_2=0.1\, {\rm meV}$.
The parameter $\Delta_Z$ models an intrinsic magnetization of the material and is varied to explore 
different phase regions, usually with $\Delta_Z< 2\, {\rm meV}$. We take these values
as reasonable estimates for QAH insulators based on  Cr doped and V doped (Bi,Sb)$_2$Se$_3$ 
or (Bi,Sb)$_2$Te$_3$ magnetic thin films \cite{Wang14,Wang15}. 
Nevertheless, 
the results we disuss below are not sensitive to small variations around these estimates.

\begin{figure}[t]
\begin{center}
\includegraphics[width=0.25\textwidth,trim=2cm 17.cm 2.5cm 6.cm,clip]{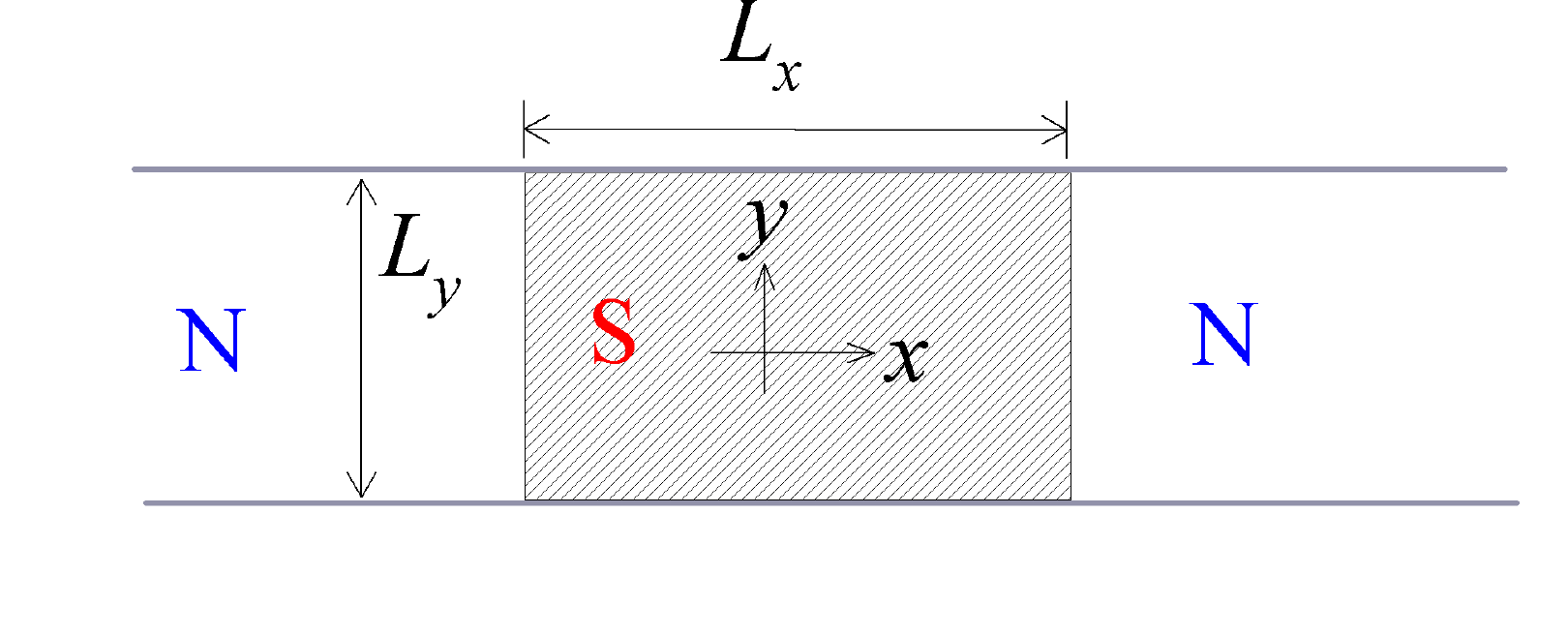}
\includegraphics[width=0.5\textwidth,clip]{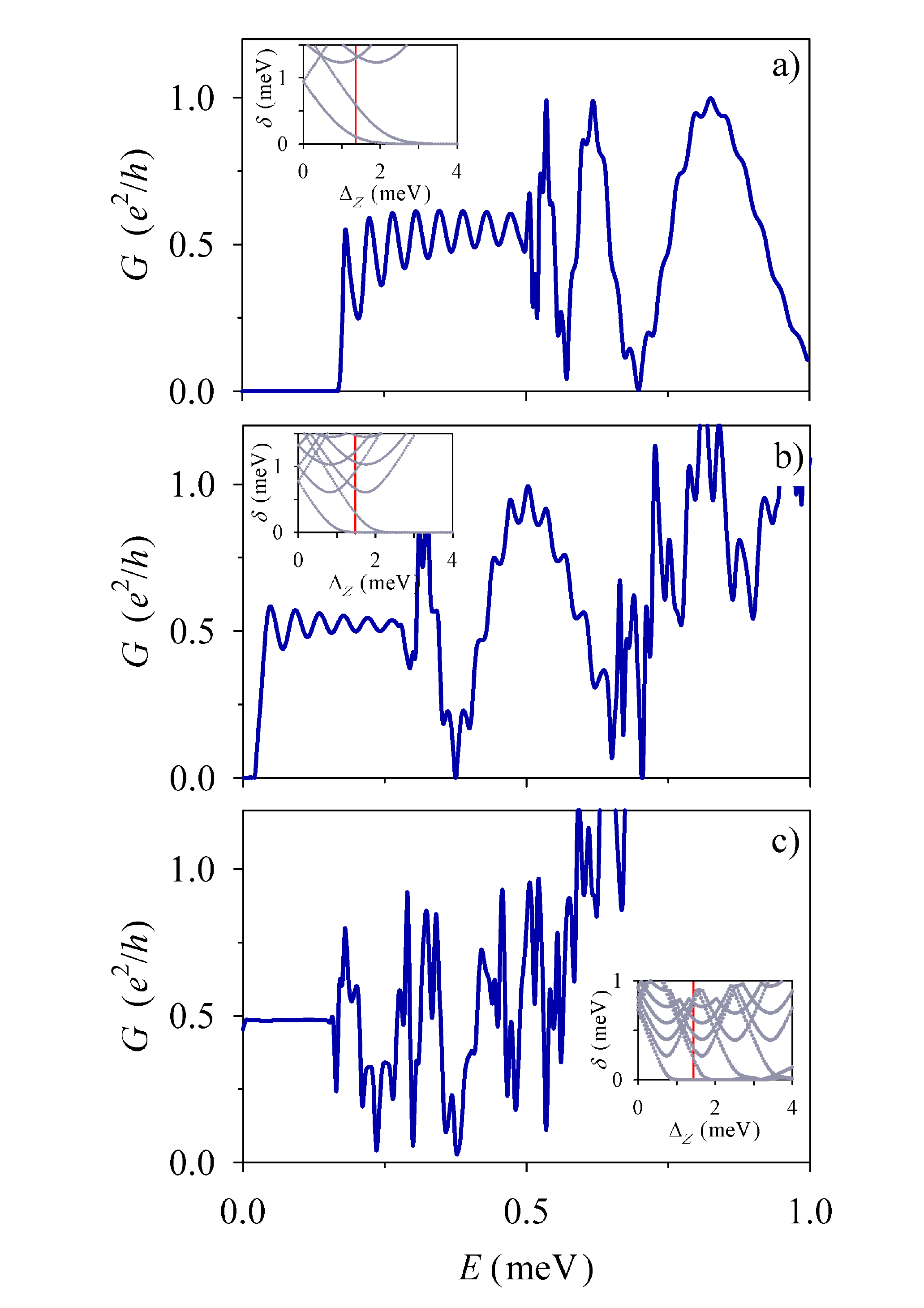}
\vspace{-3mm}
\end{center}
\caption{
Upper: Sketch of the physical system. Panels a)-c): Conductance as a function of the energy (bias)
for $L_y=1$ $\mu$m (a), 2 $\mu$m (b) and 5 $\mu$m (c).
The three panels are for $L_x=20$ $\mu$m
and $\Delta_Z=1.5$ meV.
The insets in each panel show the $\delta-\Delta_Z$ phase diagram, with $\delta$ the $k=0$ gap
of the $L_x\to\infty$ system. The vertical (red) line in each inset 
shows the correspondence with the energy sweep of its panel.}
\vspace{-5mm}
\label{Fig1}
\end{figure}

{\em Method.-} Our analysis is based on the numerical solution of the Bogoliubov-deGennes scattering equation ${\cal H}\Psi=E\Psi$ for a given energy $E$, assuming an expansion of the wave function 
in the complex band structure for each portion of the strip where the 
parameters are constant. An {\em effective} matching in 2d  at the two interfaces of the 
NSN double junction yields all the scattering coefficients. 
The method is explained in more detail in the Supplemental Material \cite{SM} and is based on Refs.\ \cite{Serra13,Osca15,Osca2017,Osca2017b,Osca18,tis01,arpack,Lent}.
In particular, the conductance for a 
given energy $E$ is determined as 
\begin{equation}
G(E) = \frac{e^2}{h} \left[ T_N(E)+ R_A(E) \right]\; .
\end{equation}
where $T_N$ is the normal (electron-electron) transmission probability and $R_A$ is the Andreev 
(electron-hole) reflection probability. The superconductor is assumed grounded and the applied bias  symmetrical since otherwise currents may emerge from the superconductor and flow to the leads \cite{Lambert,Lim12,Chung}

The  resolution method describes both longitudinal and transverse evanescent behaviour, an essential point since we are interested in the dependence  with both $L_x$ and $L_y$. Including large-enough sets of $N_k\approx 100$ complex-$k$ waves for each region describes longitudinal evanescent behaviour while a uniform $y$ spatial grid
with $N_y\approx 100$ points is required
to describe the transverse behavior.  In addition, a minimal grid of only $N_x=5$ points for each interface
is required for the matching. The computational cost is small and, quite importantly, it is independent of $L_x$, which is again essential for the study of the $L_x$-dependence up to large values.

{\em Results.-} Figure \ref{Fig1} shows a characteristic evolution 
of bias-dependent conductance, $G(E)$, as the strip width $L_y$ increases: 
1 $\mu$m (a), 2 $\mu$m (b) and  5 $\mu$m (c). As expected, the wide strip 
shows a low-energy flat conductance of 0.5$e^2/h$ that, when increasing $E$, evolves
into a complicated variation due to the successive activation of higher energy modes of the strip. The value of $\delta$, the $k=0$ gap energy, is shown for each mode 
as a funcion of $\Delta_Z$ in the corresponding insets of Fig.\ \ref{Fig1}. 
In the narrower strips of Figs.\ \ref{Fig1}a,b the flat 0.5$e^2/h$ plateau
transforms into a clear oscillating pattern, with larger 
amplitude at the onset and gradually damping with increasing energies. This 
pattern is clearly enhanced in the smaller $L_y$ strip (Fig.\ \ref{Fig1}a).
A sudden change of this pattern occurs when a second chiral
mode is activated, setting in a  larger amplitude envelope oscillation between zero and one conductance quantum in Figs.\ \ref{Fig1}a,b.

\begin{figure}[t]
\begin{center}
\includegraphics[width=0.35\textwidth,trim=1cm 14.5cm 2.5cm 2.cm,clip]{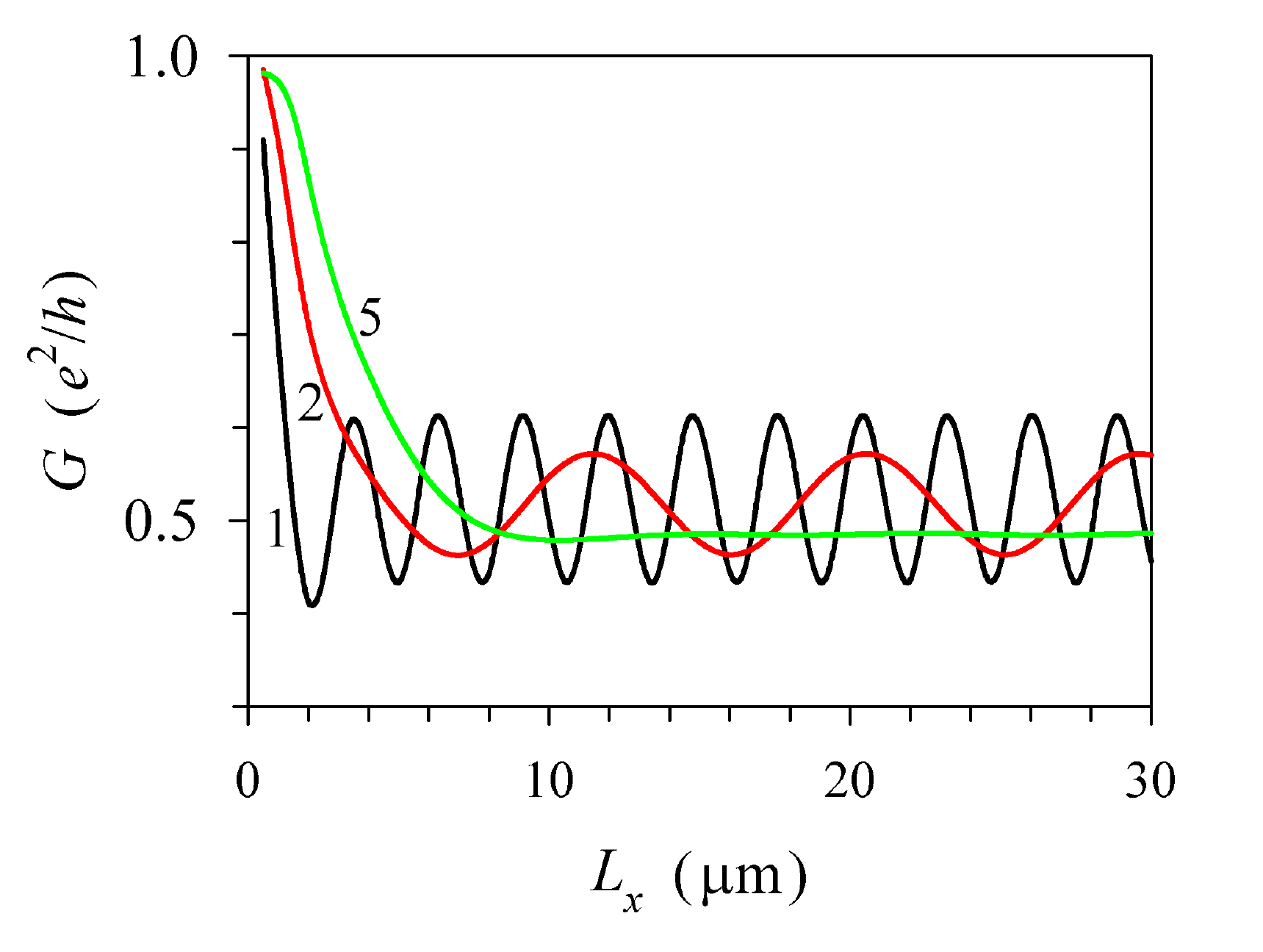}
\includegraphics[width=0.45\textwidth,trim=1cm 13.5cm 2.cm 4.cm,clip]{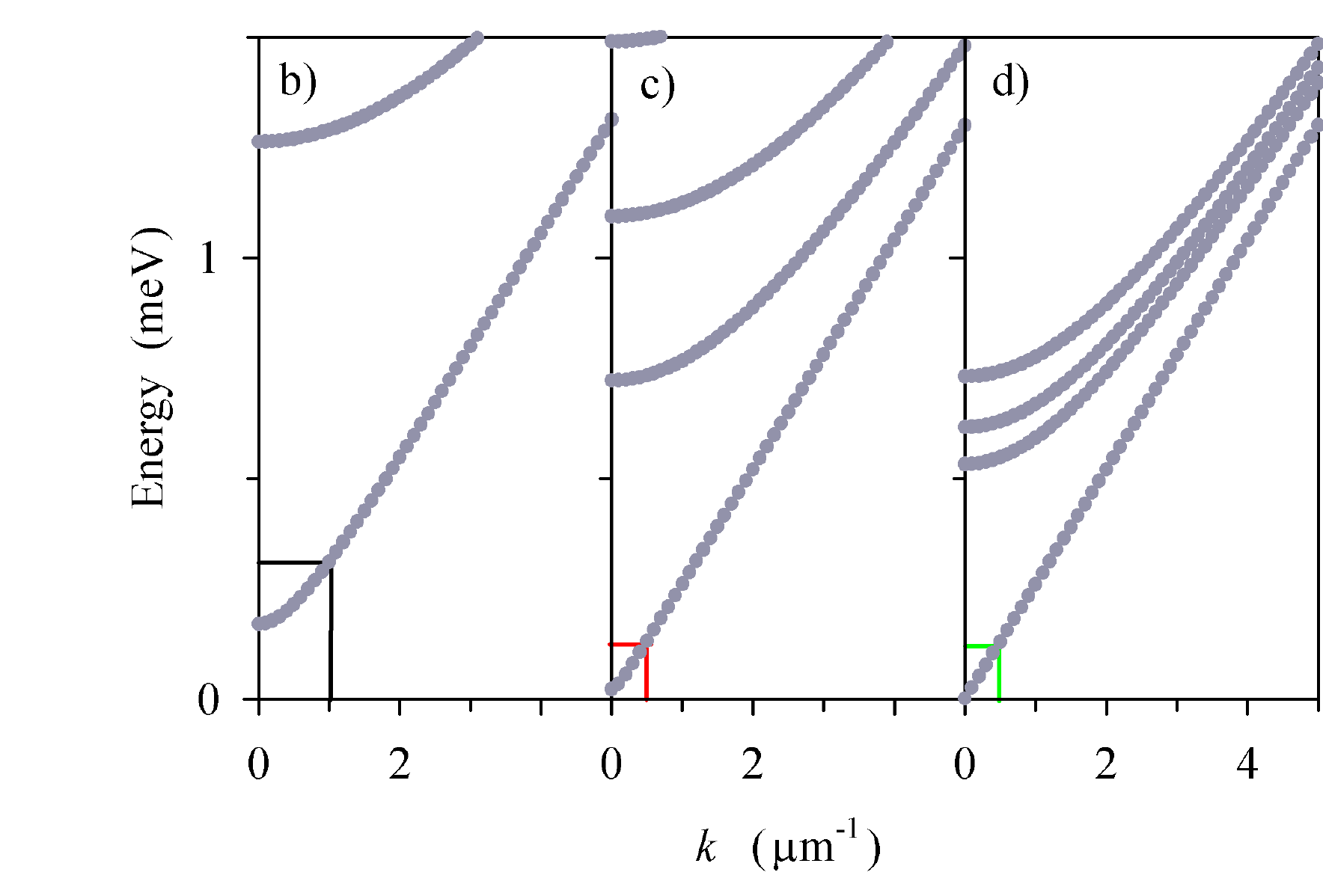}
\vspace{-3mm}
\end{center}
\caption{a) $L_x$ dependence of the conductance for different values of the width $L_y$, as indicated
by the number close to each line (in microns). The results correspond to those in Fig.\ \ref{Fig1} for selected values of the energy:
0.3 meV ($L_y=1$ $\mu$m), 0.09 meV ($L_y=2$ $\mu$m), 0.1 meV ($L_y=5$ $\mu$m). 
b)-d): Energy bands for the corresponding values of $L_y$. The energy and wavenumber are indicated in each panel with the same line color of a).
}
\vspace{-5mm}
\label{Fig2}
\end{figure}

The oscillating conductance in the regime of a single chiral mode of a narrow strip is the main result of this work. It is a sizeable effect, easily reaching 40-50 \% conductance variations for energies near the activation onset of Fig.\ \ref{Fig1}a. Next, we study the role of the longitudinal distance $L_x$ on the oscillation.
Figure \ref{Fig2} shows $G(L_x)$ corresponding to selected transverse widths 
and energies of Fig.\ \ref{Fig1}. This figure provides a quantitative measure of the 
contribution of longitudinal evanescent modes to the conductance. In all cases 
$G$ initially decreases from a value close to one, reaching a sustained regime
after a critical value of $L_x$ is exceeded. In the narrower strips the 
asymptotic-$L_x$ regime again reproduces the oscillating pattern mentioned above.

The separation between successive conductance maxima (minima) of the sustained
oscillations of Fig.\ \ref{Fig2}a can be related to the speed of the chiral Majorana mode. This connection is clear from a direct comparison between the computed 
real band structure of the propagating modes shown in Fig.\ \ref{Fig2}b-d.
The approximate Majorana mode is represented by the lowest band with an
almost linear dependence on wavenumber, $E\approx c\hbar k$, with 
$c\approx 0.26\,{\rm meV}\mu{\rm m}/\hbar$.
Assuming a conductance maximum requires an integer number of half wavelengths
fit in distance $L_x$ we infer $c=\Delta L_x E/\pi$, where $\Delta L_x$ is the 
separation of two successive maxima. This reproduces $c\approx 0.26\, {\rm meV}\mu{\rm m}/\hbar$  for the
$L_y=1\, \mu$m and $2\, \mu$m results of Fig.\ \ref{Fig2}a, respectively.

As explicitly shown above, measuring $G(L_x)$ allows an electrical determination of the 
mode speed. The same conclusion is obtained inferring $c$ from the separation 
energy $\Delta E$ between two successive maxima of the oscillation pattern 
$G(E)$
of Fig.\ \ref{Fig1}a,b. In this case, however, the oscillation is not sustained 
for arbitrary large values, but one has to choose $E$ in the proper 
interval corresponding to the propagation of a single chiral mode.
There is a clear analogy between our model system and a photon interferometer, 
similarly to other 
Condensed Matter systems such as Aharonov-Bohm rings, with the genuine 
difference that in a strip the interference of chiral modes is governed by the
interplay between strip dimensions $L_x$, $L_y$
and the mechanism of Andreev backscattering. Incidentally,
in units of the photon speed in vacuum the chiral mode speed of
Fig.\ \ref{Fig2} is $\approx 1.3\times 10^{-3}$.

\begin{figure}[t]
\begin{center}
\includegraphics[width=0.45\textwidth,trim=0cm 18cm 0cm 1.5cm,clip]{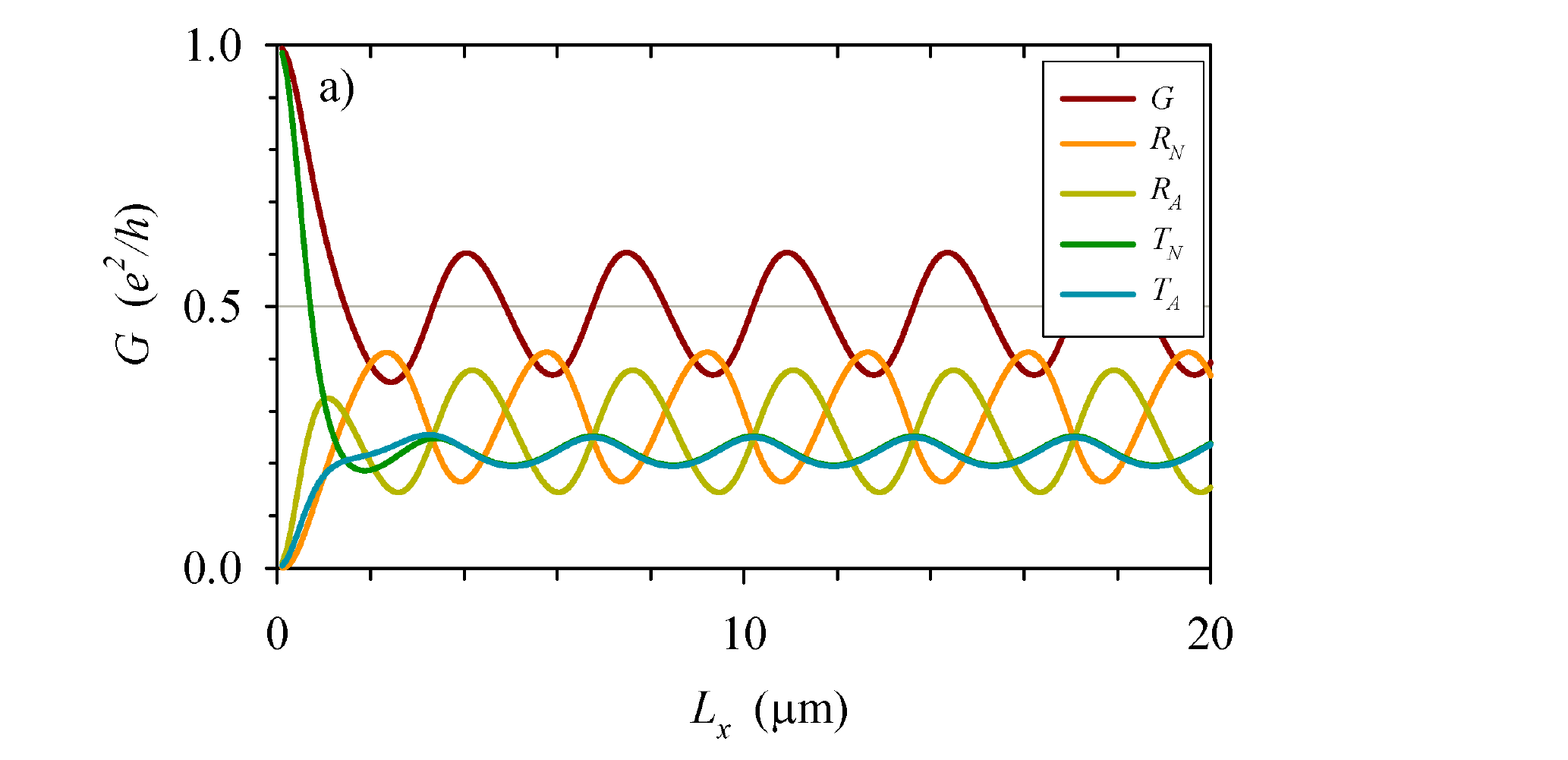}
\includegraphics[width=0.5\textwidth,trim=0cm 0cm 0cm 0cm,clip]{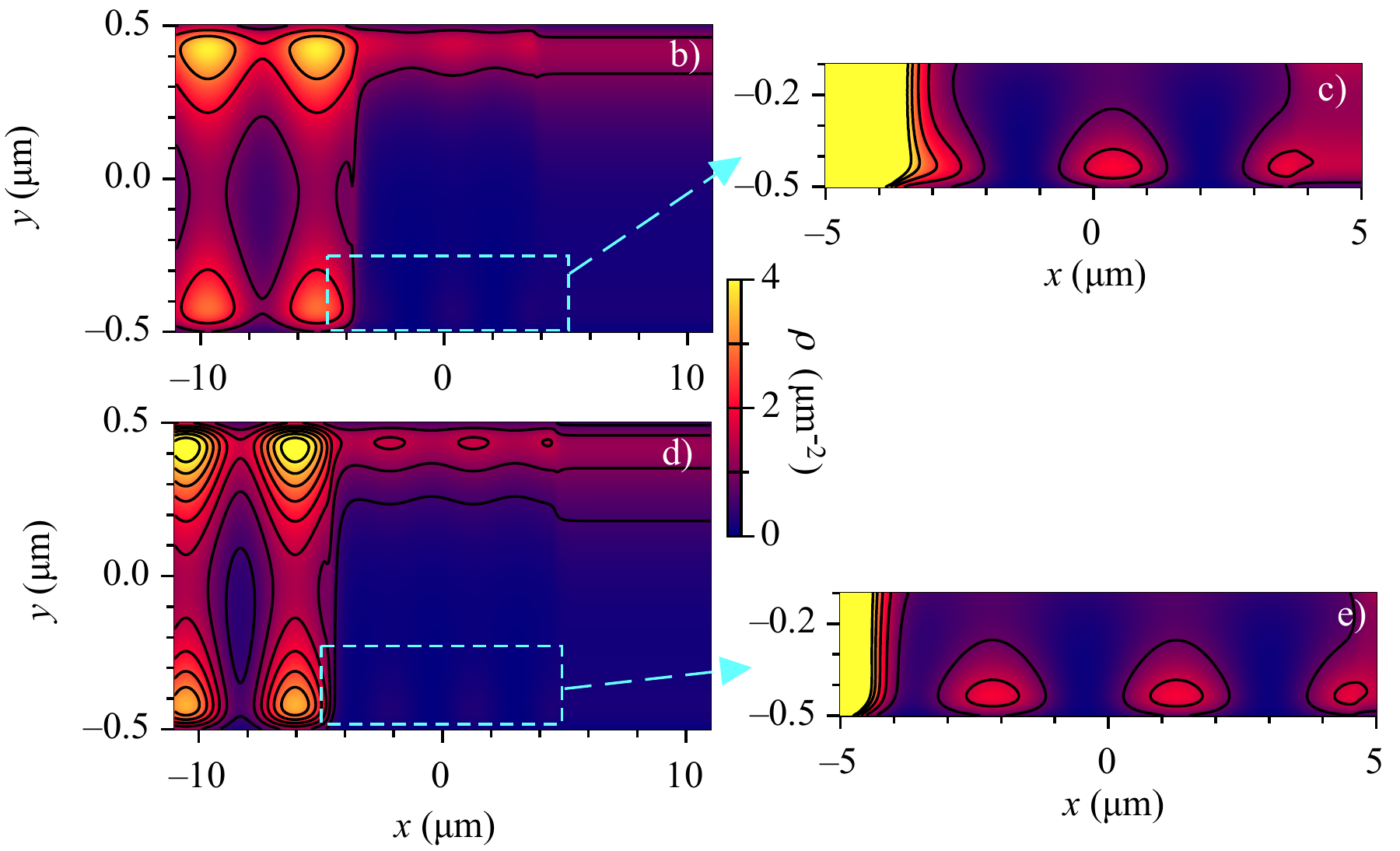}
\end{center}
\caption{
a) Evolution with $L_x$ of normal and Andreev reflections and transmissions for a strip of $L_y=1$ $\mu$m and  energy $E=0.25$ meV. b-e) Spatial distribution of probability density, in $\mu{\rm m}^{-2}$, 
corresponding to the conductance maximum at $L_x=7.5\, \mu$m (b) and minimum at $L_x=9.3\, \mu$m (d). 
Panels c) and e) show the densities in a 10x-zoomed scale for the highlighted regions.  
}
\label{Fig3}
\end{figure}

We now elucidate from our calculations the physical mechanism behind the enhanced conductance when
condition
$L_x=n\lambda/2$ is fulfilled, with $n$ an integer and $\lambda$ the wavelength of the chiral mode between the two interfaces. Figure \ref{Fig3} shows that the conductance oscillation basically reflects the 
behavior of the Andreev reflection probability $R_A$, the normal transmission $T_N$ is also oscillating
but with a weaker amplitude and a reduced wavelength. The contour plots displayed in panels b)-e)
show the distribution of quasiparticle probability density corresponding to an incidence from the upper left chiral mode. As expected, in the superconducting piece of the strip transmission proceeds predominantly attached to the upper edge. A zoom of the lower edge reveals that in panel c) two full 
intervals between density maxima 
fit in $L_x$, counting backwards from the lower right maximum, while in panel e) two and a half intervals can fit in.
As the distance between density maxima is $\lambda/2$,   
this is the interference of the backscattered chiral mode from the right interface that can thus enhance/decrease the global Andreev 
reflection from the left interface.

\begin{figure}[t]
\begin{center}
\includegraphics[width=0.45\textwidth,trim=0cm 16.5cm 0cm 1.5cm, clip]{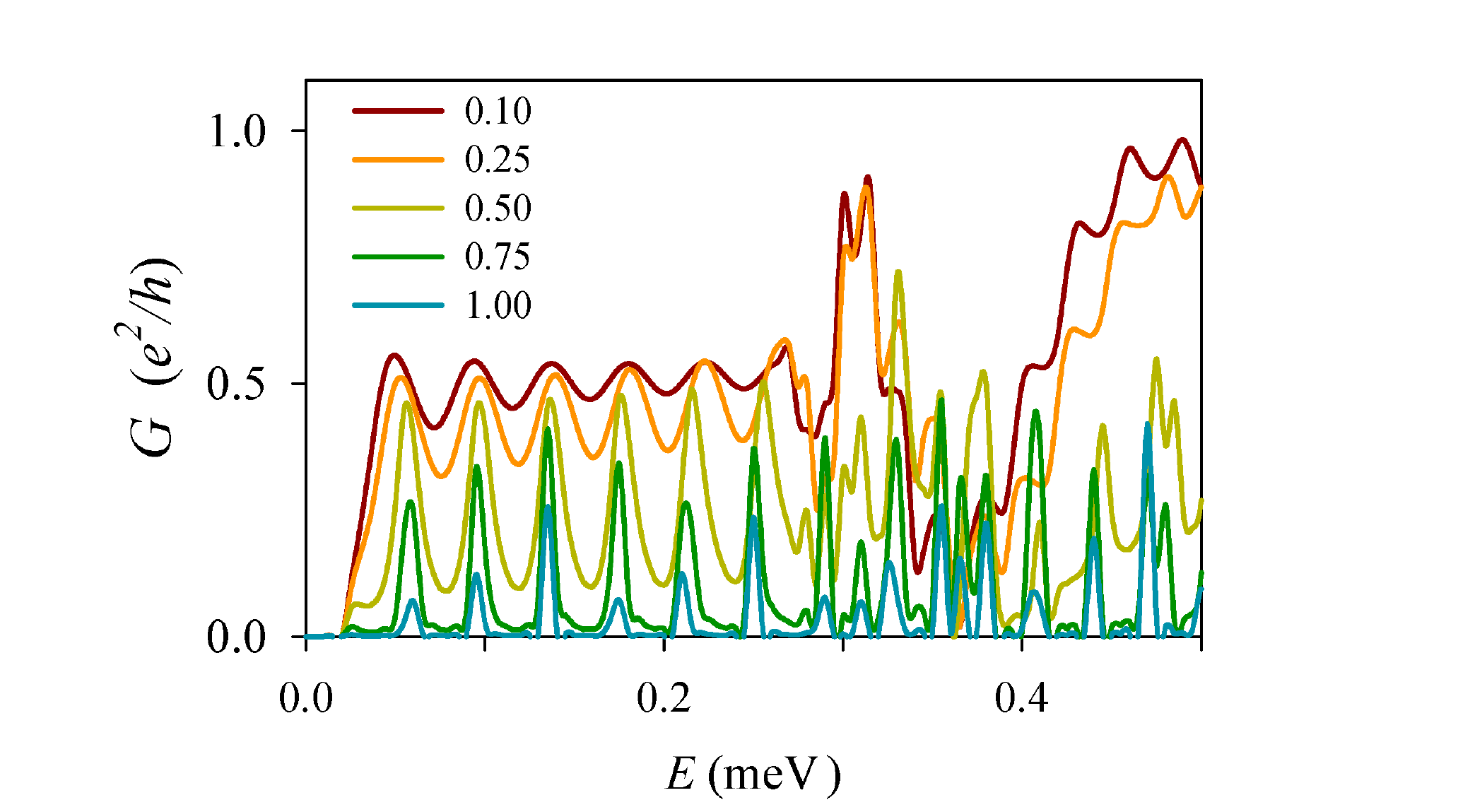}
\end{center}
\caption{$G(E)$ in presence of barriers to the left and right of the strip superconductor region. The different curves correspond to varying barrier lengths
$L_b$,
as indicated in microns. We take $m_{0b}=2$ meV in the barrier sections of the strip.
Other parameters as in Fig.\ \ref{Fig1}b.
}
\label{Fig4}
\end{figure}

Edge chiral Majorana modes are robust against backscattering induced by 
local disorder, in a similar way of the quantum Hall effect. We have checked (Supplemental Material)
that the conductance oscillations of this work are indeed robust with the presence of local fluctuations
$\delta m_0(x,y)$ in a portion of the strip modelling bulk disorder as well as deviations from perfect straight edges.
As final result, we have considered the presence of  side {\em material} barriers in the longitudinal direction 
by assuming $m_0\equiv m_{0b}=2$ meV in regions of length $L_b$ to the left and right of the 
superconductor central island of the strip. As with usual potential barriers, the transparency of those side regions can be tuned by changing $m_{0b}$ and/or $L_b$.  The presence of side barriers tends to decouple leads 
and central island, leading to a more clear manifestation of the interferometer character discussed above. Figure \ref{Fig4} shows how the oscillating pattern
of $G(E)$ evolves to a sequence of spikes as the barriers become less and less 
transparent; each spike signaling the condition of resonant backscattering of the chiral Majorana mode.
Thus, the configuration with side barriers suggests a device operation with more clear difference between its on
and off states.

{\em Conclusion.-} In summary, our calculations suggest that  a quantum-anomalous Hall thin strip with a central superconductor section could display conductance oscillations when the strip transverse size $L_y$  is in the 
micrometer range and the system is in a fully  quantum coherent regime.
As a function of the superconductor region length $L_x$ the oscillations are sustained up to arbitrary large values. We conclude that a quantum mesoscopic strip behaves as an interferometer that allows 
measuring the speed of the chiral Majorana mode. The physical mechanism is the enhanced Andreev reflection due to resonant backscattering of the chiral mode from the second to the first interface.
The presence of side barriers magnifies the interferometer effect by yielding conductance spikes.

\acknowledgements
This work was funded by MINEICO (Spain), grant MAT2017-82639.


%


\appendix

\vspace*{0.5cm}

\centerline{\bf Supplemental Material}
\vspace*{0.5cm}
We give details of the method we use to solve the Bogoliubov-deGennes equation
for quasiparticle scattering.
Some complementary results are also included.

\section{Method}

As discussed in the main text, we devise an approach to solve the scattering Bogoliubov-deGennes equation
for quasiparticles
\begin{equation}
\label{aeq1}
{\cal H}\Psi(xy\eta_\sigma\eta_\tau\eta_\lambda) = E \Psi(xy\eta_\sigma\eta_\tau\eta_\lambda)\; ,
\end{equation}
where the $\eta$'s represent the doubled variables for spin, isospin and pseudospin, respectively.
The main idea is to expand the wave function using the complex band structure for each portion 
of the wire and, subsequently, solve an effective set of equations ensuring the proper matching
of solutions in 2d at the two longitudinal interfaces. Related solution schemes have been previously used
by us in Refs. \cite{Serra13,Osca2017,Osca2017b}.

\subsection{Complex $k$'s}
In each uniform region (along $x$) we may expand
\begin{equation}
\label{aeq2}
\Psi(xy\eta_\sigma\eta_\tau\eta_\lambda)
=
\sum_{c} c_k\, e^{ik(x-x_k)}\, \Phi_k(y\eta_\sigma\eta_\tau\eta_\lambda)\; ,
\end{equation}
where the $\Phi_k$'s are the solutions of the translationally invariant problem for a characteristic wavenumber $k$. The phases $\exp{(-ikx_k)}$ are introduced for convenience by a proper choice of the $x_k$'s, to be mentioned below. It is important to include the possibility of complex wavenumbers $k$ in order to describe evanescent modes. The sets of solutions $\{k,\Phi_k\}$ are determined from 
\begin{equation}
\label{aeq3}
(h(k)-E) \Phi_k(y\eta_\sigma\eta_\tau\eta_\lambda) = 0\; ,
\end{equation}
where $h(k)$ is the reduced Hamiltonian once the $x$ dependence has been removed assuming 
a plane wave along $x$.

Equation (\ref{aeq3}) can be seen as a nonlinear eigenvalue problem for $k$ (notice that $E$ is known).
The kinetic-like term of the Hamiltonian yields a $k^2$ contribution, while the Rashba-like term 
yields a linear $k$ term. A clever transformation allows a simplification of Eq.\ (\ref{aeq3}) to a linear eigenvalue 
problem for $k$ by doubling the number of components of the wave function \cite{tis01}. 
Defining 
$\tilde\Phi_k(y\eta_\sigma\eta_\tau\eta_\lambda\eta_g)$ where $\eta_g$ is the new (generalized) double valued index, such that
\begin{equation}
\tilde\Phi_k(y\eta_\sigma\eta_\tau\eta_\lambda\eta_g) =
\left\{
\begin{array}{cc}
\rule{0cm}{0.1cm}
\Phi_k(y\eta_\sigma\eta_\tau\eta_\lambda) & {\rm if}\; \eta_g=1\;,\\
\rule{0cm}{0.1cm}
\ell_0\, k\, s(\eta_\tau) \Phi_k(y\eta_\sigma\eta_\tau\bar\eta_\lambda) & {\rm if}\;  \eta_g=2\;,
\end{array}
\right.
\end{equation}
where $\ell_0$ is our length unit and we have defined 
$\bar\eta$ to be the opposite of $\eta$ and
 $s(\eta)=\pm1$ for $\eta=1,2$, respectively.
The resulting eigenvalue equation for $\tilde\Phi_k$ reads
\begin{equation}
\label{aeq5}
\tilde{h}\,
\tilde\Phi_k(y\eta_\sigma\eta_\tau\eta_\lambda\eta_g) =
k\,
\tilde\Phi_k(y\eta_\sigma\eta_\tau\eta_\lambda\eta_g)\; ,
\end{equation}
where $\tilde{h}$ is the corresponding generalized Hamiltonian, not depending on wavenumber, and 
$k$ is now a standard (linear) eigenvalue.

Discretizing the $y$ coordinate in a uniform grid of $N_y$ points and using finite differences for the derivatives
Eq.\ (\ref{aeq5}) transforms into a matrix problem. The corresponding matrix is non Hermitian, as could be
anticipated since $k$ can be complex. We have diagonalized it using the ARPACK inverse coding routines, which allow a very efficient use of the matrix sparseness \cite{arpack}. The method yields the set of $N_k$ 
wave numbers closer to a chosen value, typically $k=0$.

\subsection{Matching in 2d}

Having obtained a large enough set of complex $k$'s and $\Phi_k$'s in each part of the strip, next step requires
matching the solutions using the apropriate boundary conditions; a process that will determine the
$c_k$'s containing the physical information of the scattering matrix. Our approach is inspired by the 
quantum-transmitting-boundary method \cite{Lent}. We introduce a minimal $x$ grid of $N_x$ points centered around  the positions of the two longitudinal interfaces. $N_x$ equals the number of points of 
the finite-difference rule for the derivatives. Typically, $N_x=3$ is enough, although we have also checked higher values $N_x=5,7,\dots$.

The algorithm has to define a closed set of equations for the unknowns
\begin{equation}
\label{aeq6}
\left\{ {c_k}'s, \Psi(xy\eta_\sigma\eta_\tau\eta_\lambda) \right\}\; ,
\end{equation}
where the $k$'s are those of output modes ($N_k$). The total number of unknowns in the unknowns set of Eq.\ (\ref{aeq6})
is $N_k+2\times 8N_xN_y$. The corresponding closed set of equations  yielding all the unknowns is summarized in Tab.\ \ref{Tab1}, where we have defined 
the projected equations
\begin{eqnarray}
\sum_{\eta_\sigma\eta_\tau\eta_\lambda}\int{dy}\,
\Phi_{k'}^*(y\eta_\sigma\eta_\tau\eta_\lambda)\,
\Psi(xy\eta_\sigma\eta_\tau\eta_\lambda)
&=& \nonumber\\
\label{aeq7}
 \sum_k{c_k e^{ik(x-x_k)}{\cal M}_{k'k}}\; ,&& 
\end{eqnarray}
with the mode-overlap matrix
\begin{equation}
{\cal M}_{k'k}
=
\sum_{\eta_\sigma\eta_\tau\eta_\lambda}\int{dy}\,
\Phi_{k'}^*(y\eta_\sigma\eta_\tau\eta_\lambda)\,
\Phi_{k}(y\eta_\sigma\eta_\tau\eta_\lambda)\; .
\end{equation}

The resulting set of linear equations from Tab.\ \ref{Tab1} is solved using again the ARPACK 
routines. This has to be repeated for each input mode in order to determine the full scattering
probabilities for arbitrary incidence. In each case, the solutions fulfil conservation of quasiparticle probability. Since the $x$-grids are defined only around the interfaces, the computational burden is not
increasing with $L_x$. Finally, the $x_k$'s in Eq.\ (\ref{aeq2}) are chosen such that for complex 
wavenumbers
the exponentially increasing solutions always remain bounded. 

\begin{table}[t]
\begin{center}
\caption{\label{Tab1}Summary of effective 2d matching equations}
\begin{ruledtabular}
\begin{tabular}{l|l}
 & Number of Eqs.\\
 \hline
 \rule[-0.45cm]{0cm}{1cm}
 \parbox{5.5cm}{Discretized Eq.\ (\ref{aeq1}) on central points of $x$-derivative rule\\ } & $2\times8N_y$ \\
 \hline
 \rule[-0.45cm]{0cm}{1cm}
\parbox{5.5cm}{Eq.\ (\ref{aeq2}) on side points of $x$-derivative rule} & $2\times8(N_x-1)N_y$ \\
\hline
 \rule[-0.45cm]{0cm}{1cm}
\parbox{5.5cm}{Projected Eq.\ (\ref{aeq7}) on central points of $x$-derivative rule} & $N_k$ \end{tabular}
\end{ruledtabular}
\end{center}
\end{table}

\section{Results}

The following pages contain  additional results not included in the main text.
\vspace*{0.5cm}

\acknowledgements
This work was funded by MINEICO (Spain), grant MAT2017-82639.

\begin{figure*}
\begin{center}
\includegraphics[width=0.55\textwidth,trim=0cm 0cm 0cm 0cm, clip]{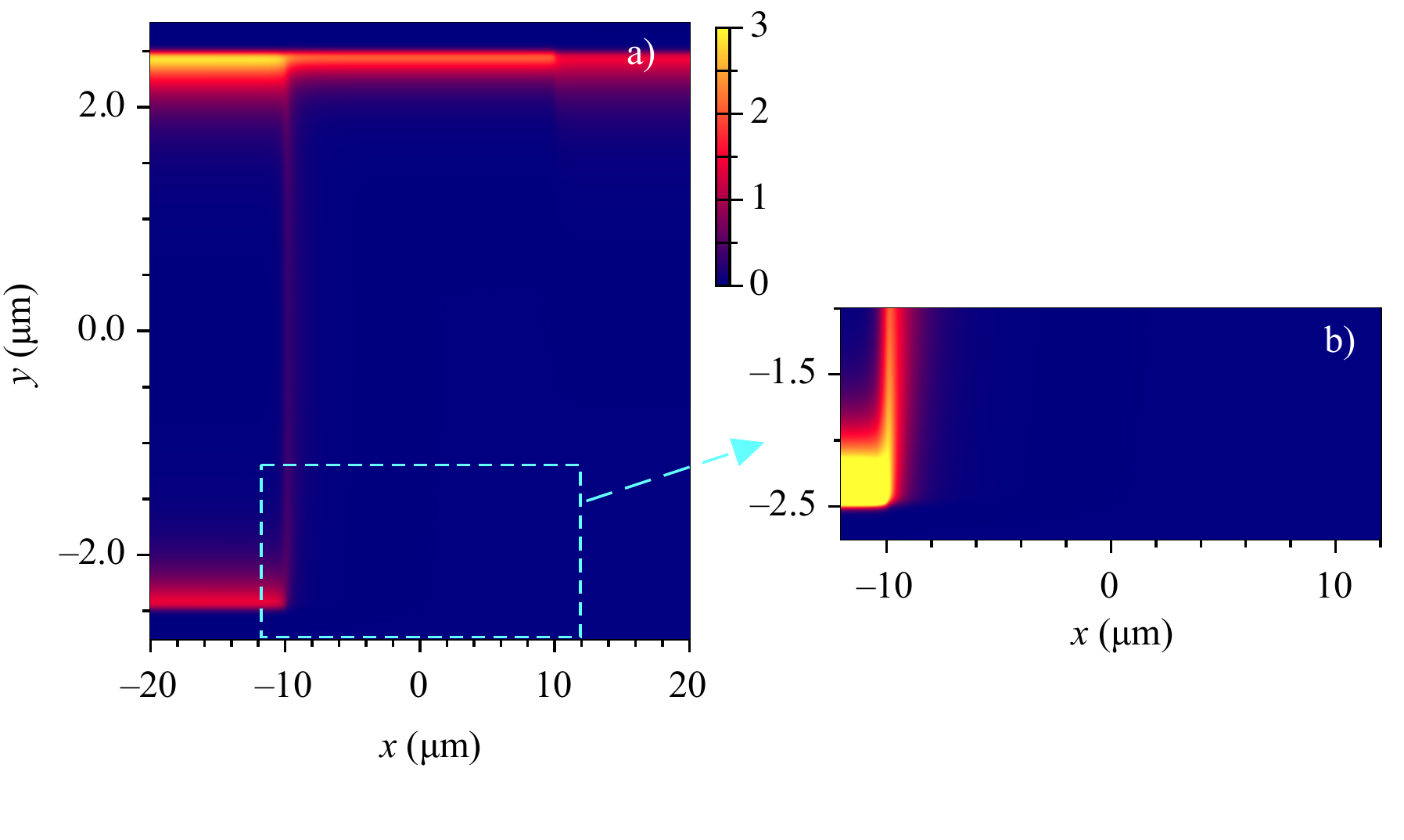}
\end{center}
\caption{Same as Fig.\ 3d,e of the main text but for a wide strip of $L_y=5\; \mu$m, $L_x=20\, \mu$m and $E=0.1$ meV. The zoomed scale shows that chiral mode backscattering has fully vanished
in between the two interfaces. Notice also that, relative to the tranverse size $L_y$, the edge-character of the modes is much increased while oscillations along $x$ are absent. We have used a $y$ grid with $N_y=200$ points. This case already agrees with the macroscopic result, with a flat conductance $G=0.5\, e^2/h$.}
\label{Fig1SM}
\end{figure*}

\begin{figure*}
\begin{center}
\includegraphics[width=0.65\textwidth,trim=0cm 16cm 0cm 2cm, clip]{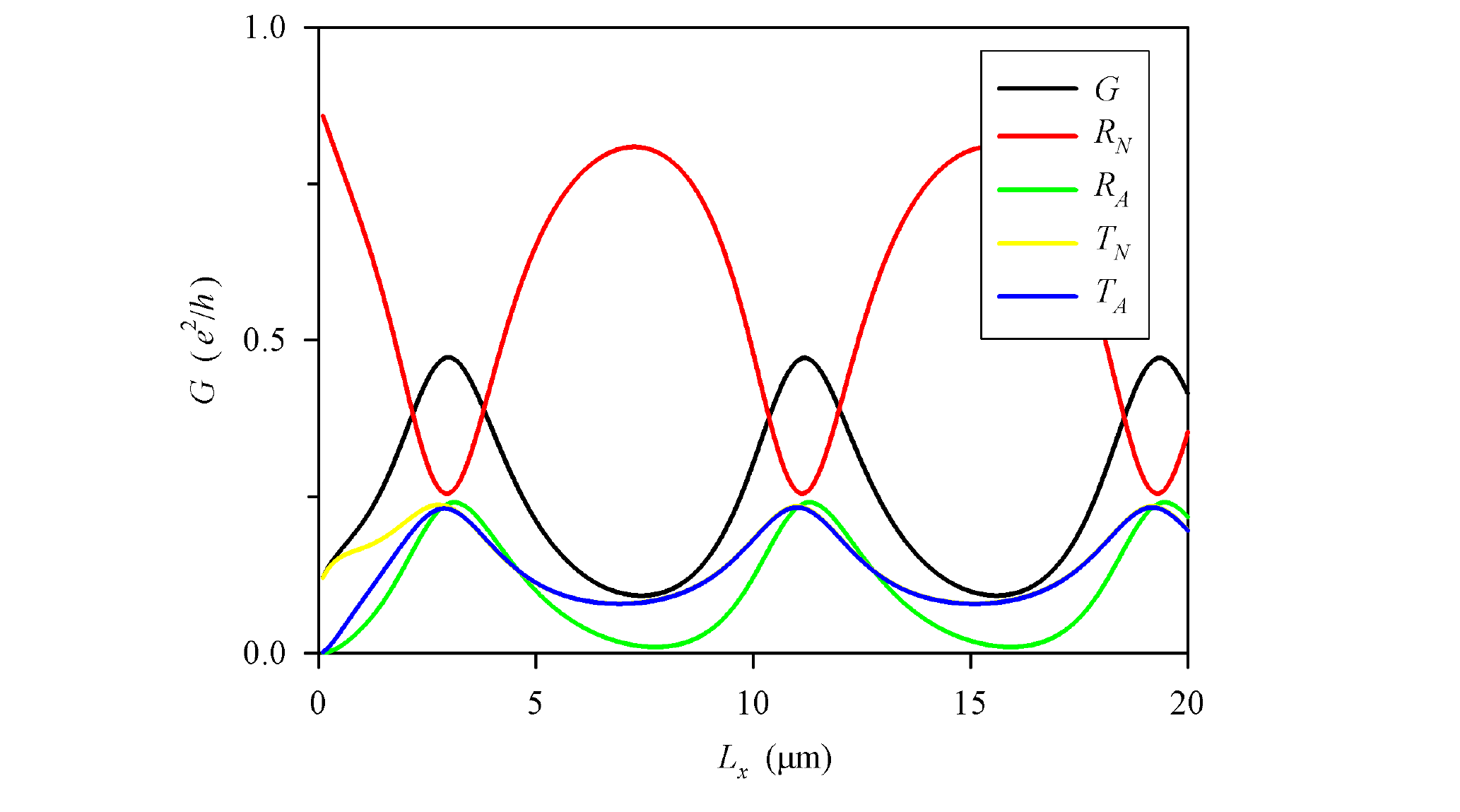}
\end{center}
\caption{$G(L_x)$ in presence of lateral material barriers with $m_{0b}=2\, {\rm meV}$ and $L_b=0.5\, \mu{\rm m}$. We have assumed $L_y=2\,\mu{\rm m}$, $E=0.1\, {\rm meV}$ and $\Delta_Z=1.5\, {\rm meV}$.
As discussed in the main text with the case of $G(E)$, due to the barriers the conductance oscillations
evolve to a sequence of peaks also with $G(L_x)$.}
\label{Fig2SM}
\end{figure*}

\begin{figure*}
\begin{center}
\includegraphics[width=0.55\textwidth,trim=0cm 0cm 0cm 0cm, clip]{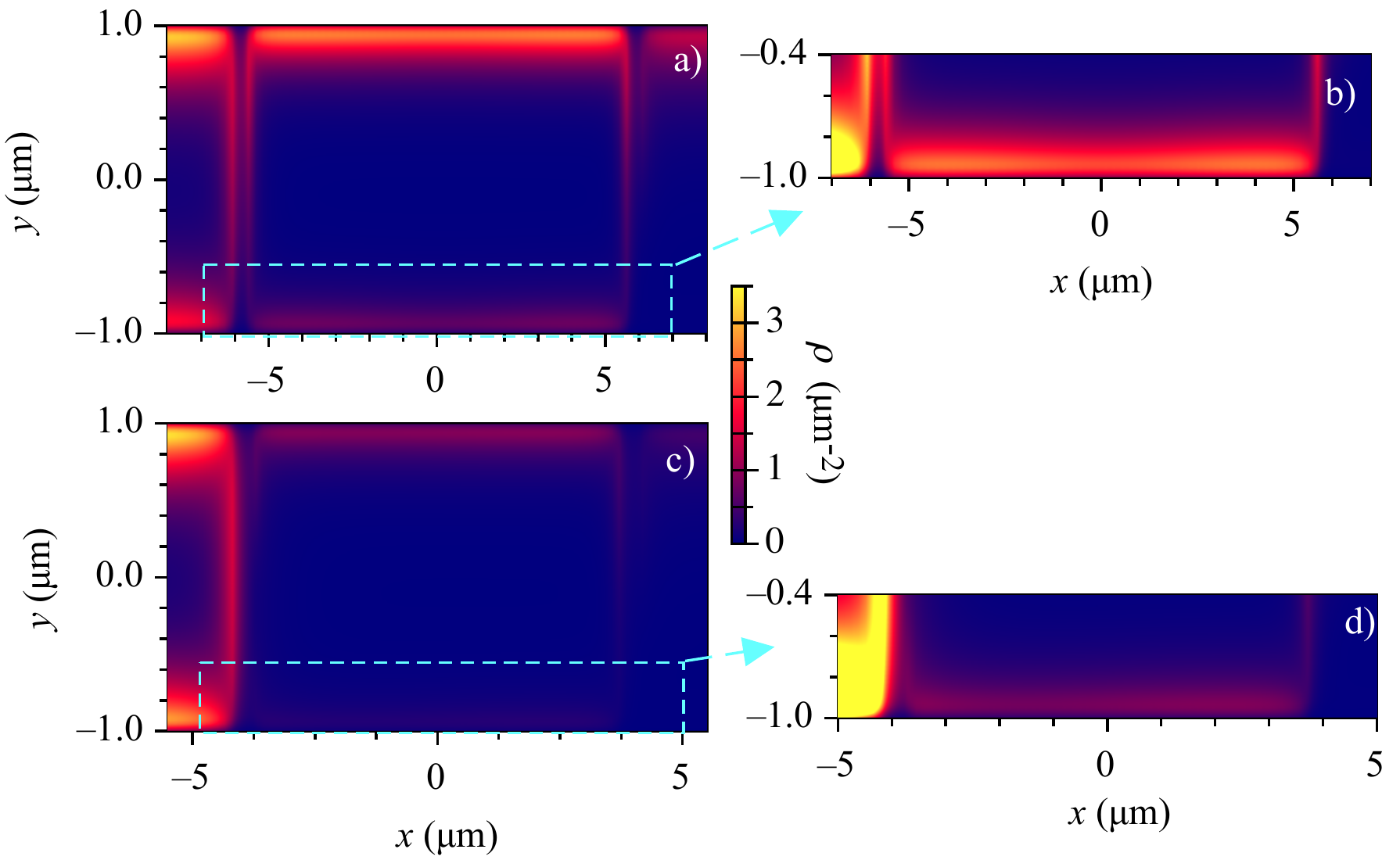}
\end{center}
\caption{Same as Fig.\ 3 of the main text but for a strip of $L_y=2\; \mu$m with side barriers of $m_{0b}=2$ meV and $L_b=0.5\, \mu{\rm m}$.
Panels a,b are for a conductance maximum (Fig.\ \ref{Fig2SM}) with
 $L_x=11.2\, \mu$m, while panels c,d are for a conductance minimum (Fig.\ \ref{Fig2SM}) with 
 $L_x=7.4\, \mu$m.
 The zoom in density scale of panels b,d is 3x.
 As expected, the backscattered mode from the second to the first interface is much weaker in
 panel d) than in b). 
 }
\label{Fig3SM}
\end{figure*}

\begin{figure*}
\begin{center}
\includegraphics[width=0.45\textwidth,trim=0cm 3cm 0cm 0cm, clip]{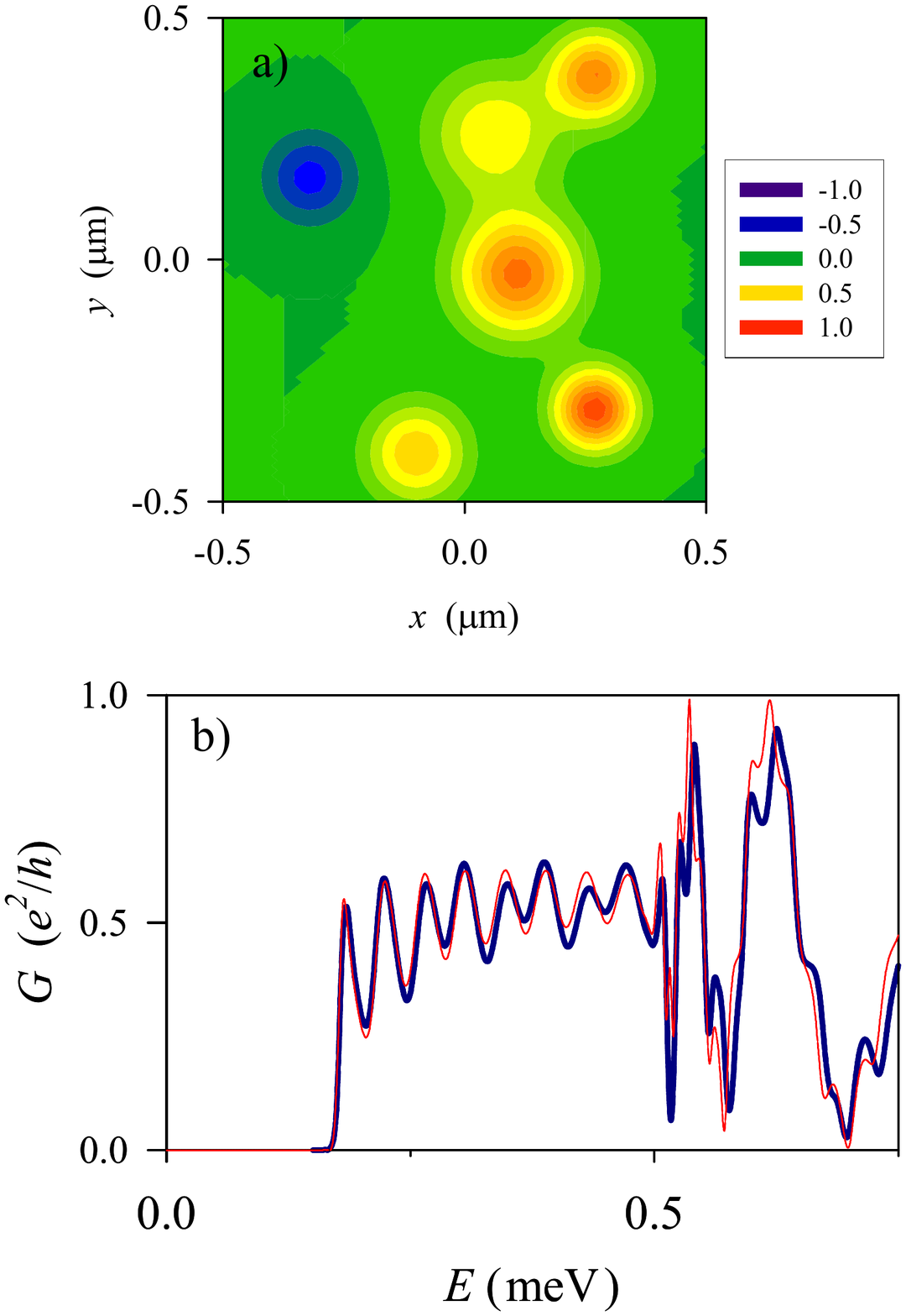}
\end{center}
\caption{Same result of Fig.\ 1a of the main text in presence of impurities modeled as a fluctuating field
$\delta m_0(x,y)$. We assumed a fluctuation range $\delta m_0/m_0\in [-1,1]$.
a) Spatial distribution of $\delta m_0(x,y)$; b) comparison of the conductance in presence (thick) and absence (thin) of the impurities.}
\label{Fig4SM}
\end{figure*}

\begin{figure*}
\begin{center}
\includegraphics[width=0.45\textwidth,trim=0cm 3cm 0cm 0cm, clip]{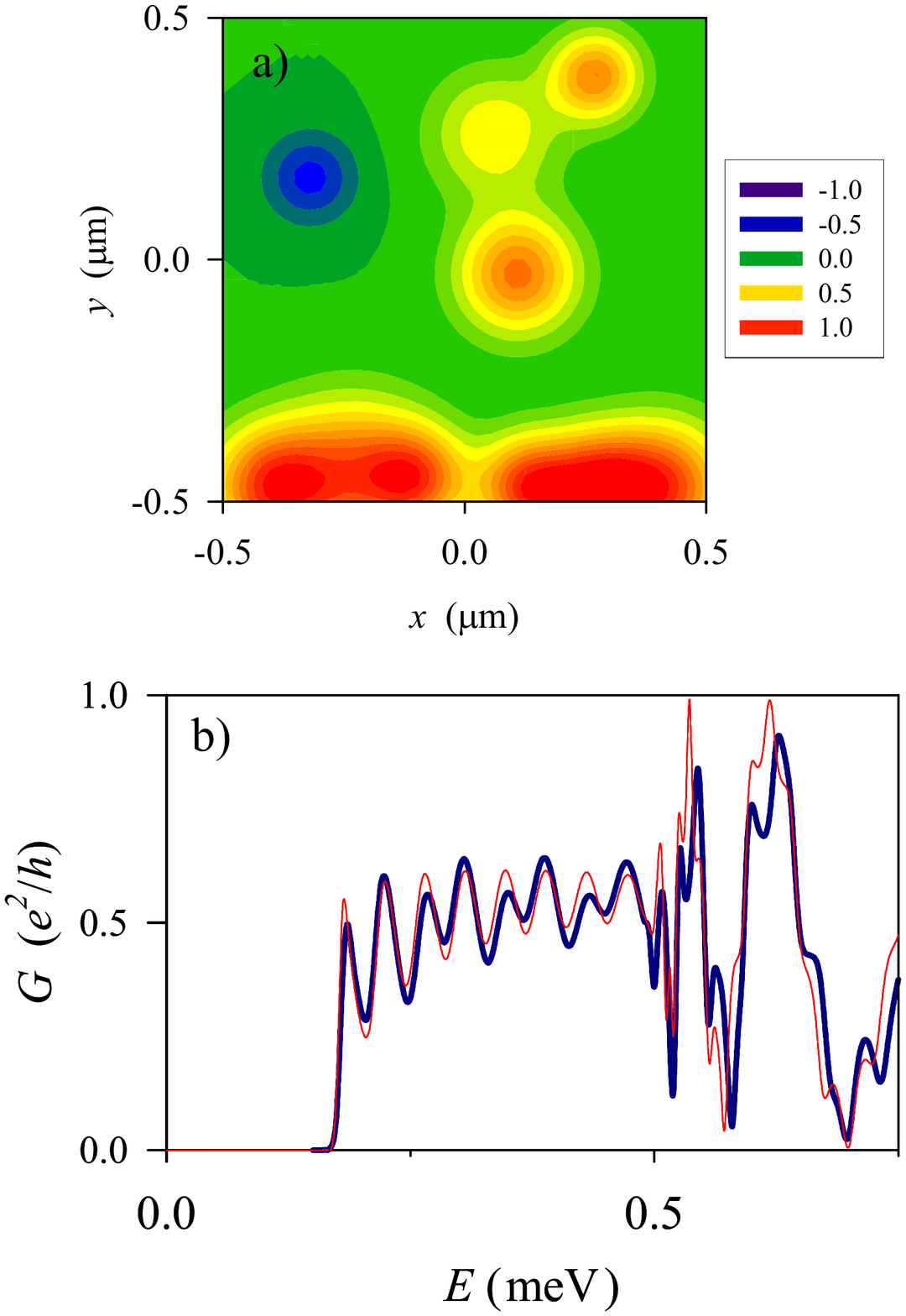}
\end{center}
\caption{Same as Fig\ \ref{Fig4SM} with an extra accumulation of impurities on the lower
edge of the strip that models a perturbation of the straight edge. }
\label{Fig5SM}
\end{figure*}

\end{document}